\documentclass[aps,prd,tightenlines,nofootinbib,reprint,superscriptaddress]{revtex4-1}

\usepackage{graphicx}
\usepackage{dcolumn}
\usepackage{bm}
\usepackage{amsmath}
\usepackage{multirow}

\begin{document}

\newcommand{\mevcc}{\!\mathrm{MeV}\!/c^2}
\newcommand{\mevc}{\!\mathrm{MeV}/\!c}
\newcommand{\mev}{\!\mathrm{MeV}}
\newcommand{\gevcc}{\!\mathrm{GeV}/\!c^2}
\newcommand{\gevc}{\!\mathrm{GeV}/\!c}
\newcommand{\gev}{\!\mathrm{GeV}}

\title{Measurements of $\Omega^-$ Branching Fractions}
\author{T.~Xiao}
\affiliation{Northwestern University, Evanston, Illinois 60208, USA}
\author{S.~Dobbs}
\affiliation{Florida State University, Tallahassee, Florida 32306, USA}
\author{A.~Tomaradze}
\author{Kamal~K.~Seth}
\affiliation{Northwestern University, Evanston, Illinois 60208, USA}


\date{\today}

\begin{abstract}
Using $e^+e^-$ annihilation data taken at the CESR collider with the CLEO-c detector, measurements of two main $\Omega^-$ branching fractions have been made using the reaction $\psi(2S)\to\Omega^-\overline{\Omega}^+$, hyperon pair production at $\sqrt{s}=3.69$~GeV, the peak of the $\psi(2S)$ resonance. $\Omega^-$ decay channels are identified through momentum distributions of charged particles, and systematics of the  $\Omega^-$ branching fractions have been studied.
The result gives:
$\mathcal{B}(\Omega^- \to \Lambda^0 K^-) = (68.9\pm9.5\pm4.3)\%$, and $ \mathcal{B}(\Omega^- \to \Xi^0 \pi^-) = (19.0\pm4.0\pm1.2)\%$.

\end{abstract}

\maketitle



\section{Introduction}

Experimental data for $\Omega^-$ hyperon ($S=-3$) are much more sparse than for $\Lambda$ and $\Sigma$ ($S=-1$), and $\Xi$ ($S=-2$) hyperons, and $\Omega^-$ decay branching fractions have only been measured once before with the hyperon beam constructed at the CERN SPS, first in 1979~\cite{omega1,omega2}, and later at higher energy in 1984~\cite{omega}. In the 1984 measurement 16,000 $\Omega^-$ were identified, and the branching fractions $\Omega^- \to \Lambda^0 K^-$, $\Omega^- \to \Xi^0 \pi^-$, $\Omega^- \to \Xi^- \pi^0$ were reported. In a recent paper~\cite{hyperon} we have reported the first measurements of $\Lambda$, $\Sigma^{+0}$, $\Xi^{+0}$, and $\Omega^-$ hyperons in their production in $e^+e^-$ annihilation at CLEO. Although compared to the hyperon beam measurement of CERN we identified far fewer $\Omega^-$ ($N=326\pm19$), since no other measurements of the branching fractions of $\Omega^-$ have been reported, we consider it important to report the results of our independent measurements of the branching fractions for the two body decays of $\Omega^-$ to $\Lambda^0 K^-$ and $\Xi^0 \pi^-$.

\section{Data Samples and Event Selections}

We use data taken with the CLEO-c detector, which has been described in detail elsewhere~\cite{cleodetector}.
The data were taken at $\psi(3686)$, $\sqrt{s}=3.69$~GeV, with integrated luminosity of $\mathcal{L}=48$~pb$^{-1}$.

We identify the $\Omega^-$ hyperon by its principal decay mode~\cite{pdg}: $\Omega^- \to \Lambda^0 K^-$ (67.8\%) [charge conjugate decay modes are included].
The event selections used to reconstruct $\Omega^-$ hyperon decays are similar to those described in our previous publication~\cite{hyperon}, and are briefly described below.

Charged particles ($\pi^\pm$, $K^\pm$, $p/\bar{p}$) are required to have $|\cos\theta|<0.93$, where $\theta$ is the polar angle with respect to the $e^+$ beam.  To identify charged particles, we use the combined likelihood variable
\begin{small}
$$\Delta \mathcal{L}_{i,j} = [-2\ln L^\mathrm{RICH} + (\chi^{dE/dx})^2]_i - [-2\ln L^\mathrm{RICH} + (\chi^{dE/dx})^2]_j,$$
\end{small}
where $i,j$ are the particle hypotheses $\pi,K,p$. The measured energy loss in the drift chamber is $dE/dx$, and $L^\mathrm{RICH}$ is the log-likelihood of the particle hypothesis using information from the RICH detector.
We identify kaons by requiring that the measured properties of the charged particle be more like a kaon than either a charged pion or proton by $3\sigma$, i.e., $\Delta \mathcal{L}_{K,\pi}<-9$ and $\Delta \mathcal{L}_{K,p}<-9$.  


We identify hyperons by kinematically fitting them under the assumption that all particles originate from a common vertex, and require that this vertex be displaced from the interaction point by $>3\sigma$.  
The $\Lambda^0$ hyperons are reconstructed by combining two oppositely charged tracks.  The higher momentum track is required to be identified as a proton, and the lower momentum track is assumed to be a negative pion.  
Each $\Lambda^0$ candidate is further required to be consistent with its nominal mass of $M(\Lambda^0)=1115.683$~MeV~\cite{pdg} within $5\sigma$. It is then kinematically fit to this nominal mass, and is required to have a decay vertex at a greater distance from the interaction point than that of the $\Omega^-$ decaying into $\Lambda^0$. The $\Omega^-$ hyperons are reconstructed by combining a $\Lambda^0$ candidate with a charged track identified as $K^-$. 

\begin{figure}[!tb]
\begin{center}
\includegraphics[width=3.5in]{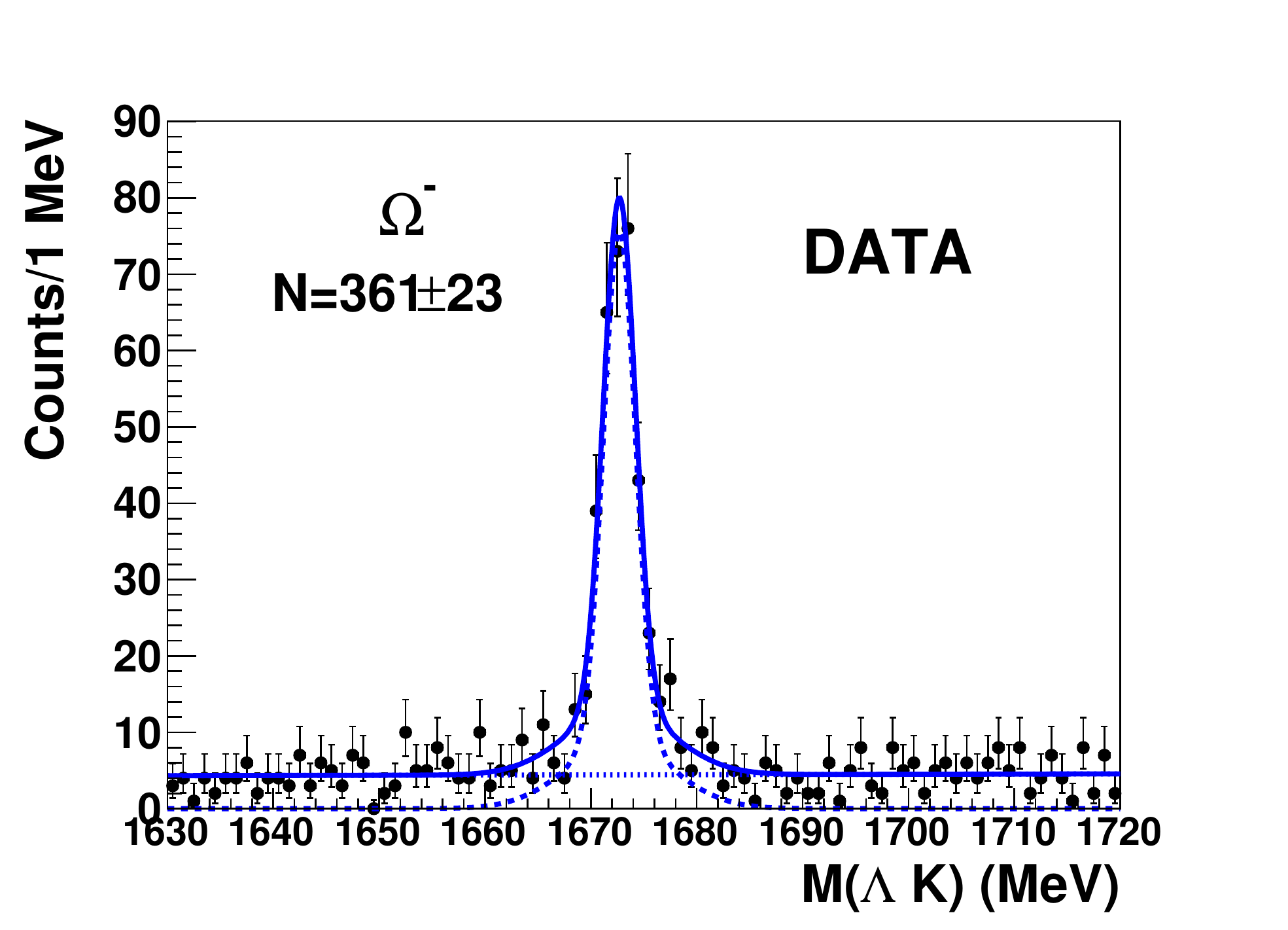}
\end{center}
\caption{ Invariant mass distributions for $\Omega^-$ candidate in $\psi(2S)$ data in the pair-production region given by $E(B)/E(\text{beam})=0.99-1.01$.}
\label{fig:omega_mass}
\end{figure}

The yield of the pair-produced $\Omega^-$ hyperons can be conveniently obtained as the events which satisfy the requirement $[E(\Omega^-)~\text{or}~E(\overline{\Omega^-})]/E(\text{beam})=0.99-1.01$.  The invariant mass distributions of these events is shown in Fig.~\ref{fig:omega_mass}. A double Gaussian signal and a constant background is used to fit the spectrum. The number of $\Omega^-$ from the fit is  $N_\Omega = N_\text{fit} - N_\text{ff}=(361\pm23)-(1\pm1)=360\pm23$, where $N_\text{ff}$ is the contribution of form factor events in these data, which are estimated by pQCD-based extrapolations, assuming a $s^{-5}$ cross section dependence, from the timelike form factor measured at $\psi(3770)$ in~\cite{hyperon}.


After reconstructing one $\Omega^-$ or $\overline{\Omega}^+$ hyperon, we can measure its branching fractions by studying the rest of tracks which decay from the other pair-produced $\overline{\Omega}^+$ or $\Omega^-$ hyperon.  
The rest of the tracks must meet the following requirements:
\begin{enumerate}
\item The track should be more like a pion or a kaon than a proton, i.e., $\Delta \mathcal{L}_{\pi,p}<0$ or $\Delta \mathcal{L}_{K,p}<0$.
\item It is required that the track should not form a $\Lambda^0$ combining another track. As before, we identify the $\Lambda^0$ by kinematically fitting two oppositely charged tracks, assuming that one is a $\pi^-/\pi^+$ and the other is identified as a $p/\bar{p}$ , to its nominal mass, $M(\Lambda^0)=1115.683$~MeV~\cite{pdg} within $5\sigma$. 
The decay vertex of $\Lambda^0$ is further required to be displaced from the interaction point by $2\sigma$.
\end{enumerate}

\section{Results}

\begin{figure}[h]
\begin{center}
\includegraphics[width=3.5in]{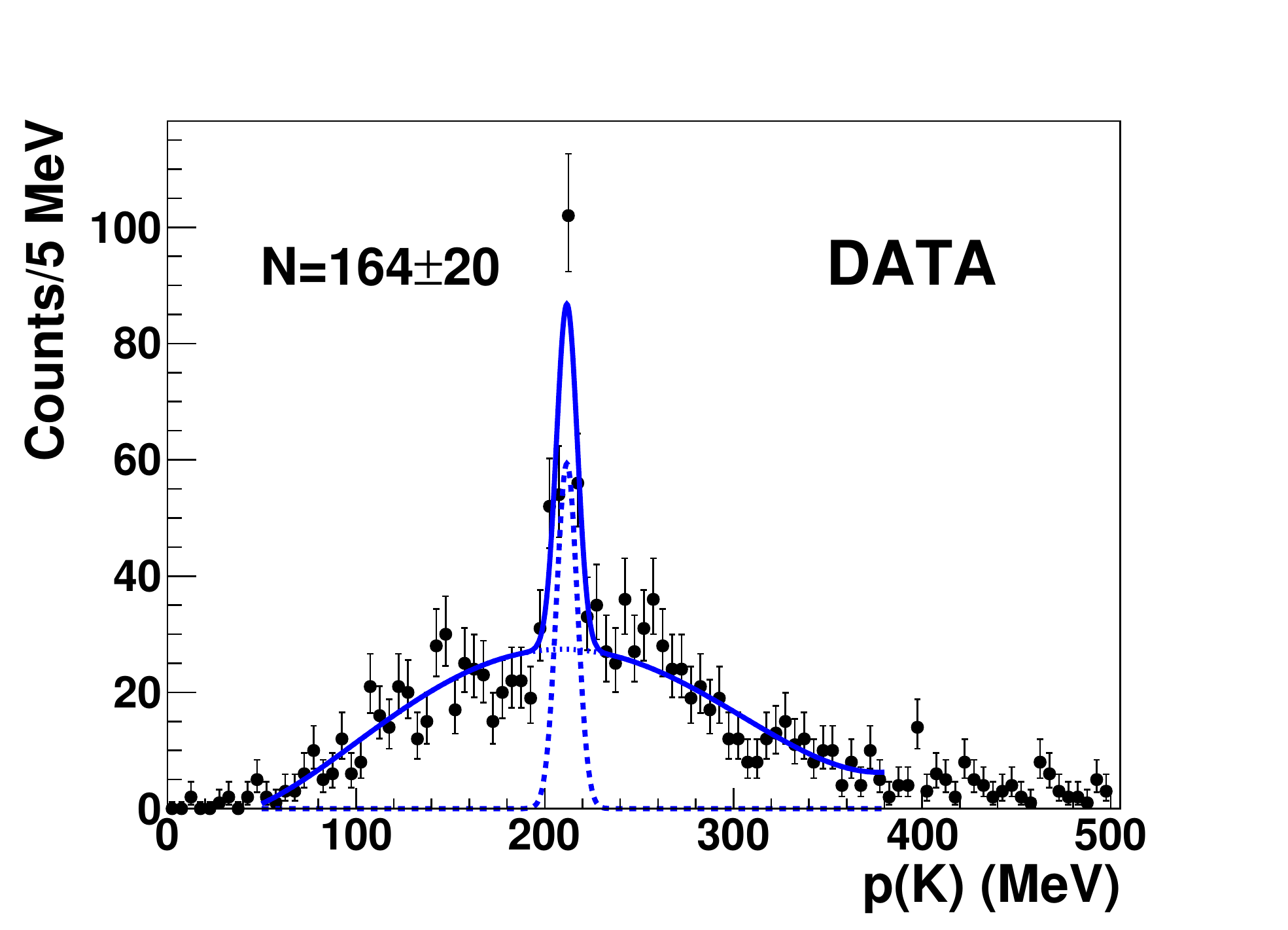}
\includegraphics[width=3.5in]{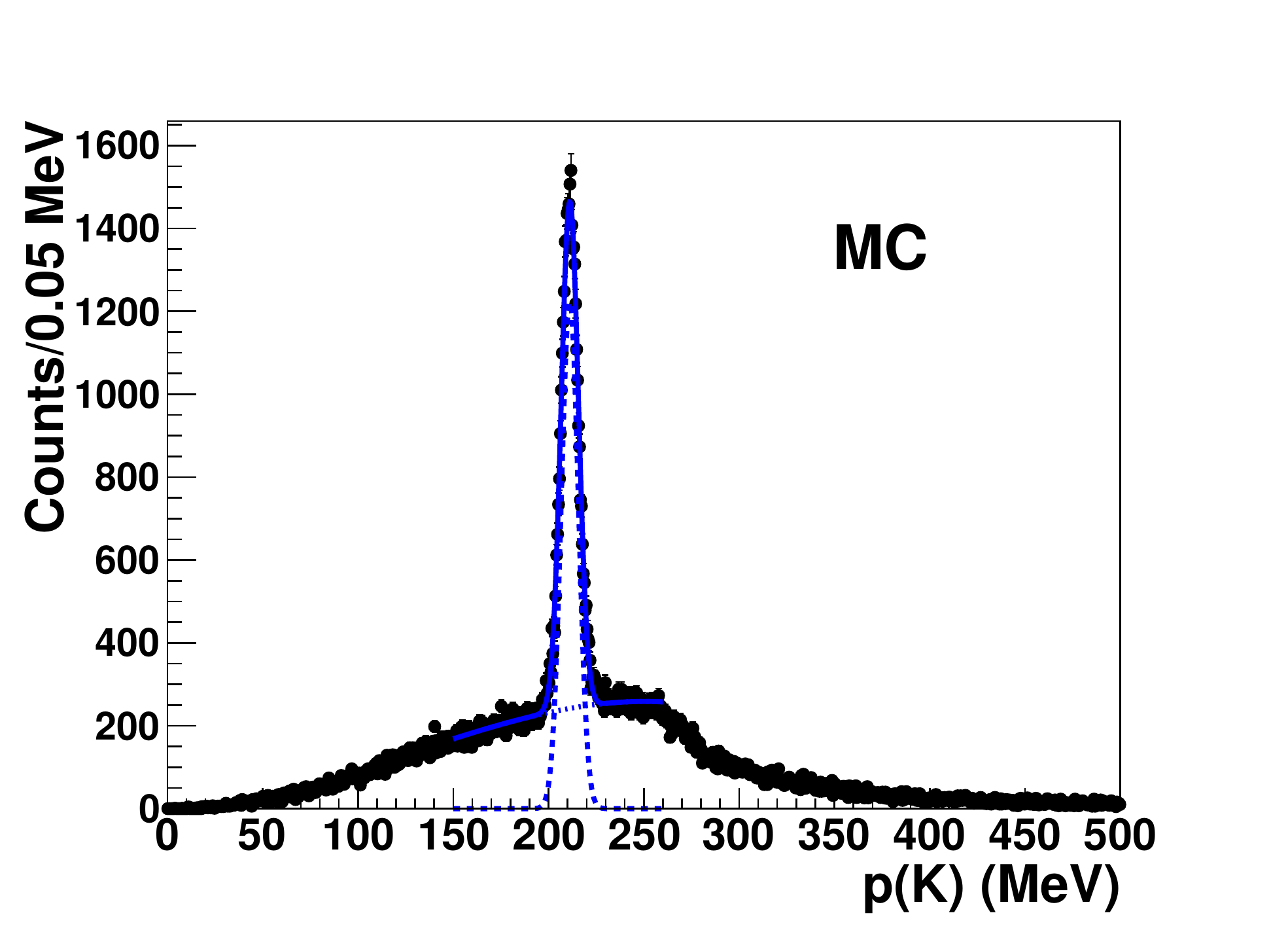}
\end{center}
\caption{Momentum distribution of tracks from $\Omega^-$ decays in the $\psi(2S)$ data (top) and MC (bottom), assuming they are kaons.}
\label{fig:k_mom}
\end{figure}

The momentum distribution of the selected tracks are calculated in the rest frame of $\Omega^-$ for the respective decays. 
Fig.~\ref{fig:k_mom} shows the momentum distribution of the tracks in the rest frame of $\Omega^-$, assuming they are kaons. It is fit with a Gaussian signal and a 3rd order polynomial background as is found to fit the Monte Carlo (MC) simulation well. The number of kaons from the fit to the momentum distribution is $N_K = 164\pm20$.

The branching fraction for $\Omega^- \to \Lambda^0 K^-$ is calculated as
\begin{equation}
\mathcal{B}(\Omega^- \to \Lambda^0 K^-) = \frac{ N_K }{ \epsilon_K\, N_\Omega} ,
\end{equation}
where $\epsilon_K=66.1\%$ is the MC-determined efficiency. 
Thus, $\mathcal{B}(\Omega^- \to \Lambda^0 K^-) = (164\pm20)/(0.661\times(360\pm23))=(68.9\pm9.5)\%$.

\begin{figure}[h]
\begin{center}
\includegraphics[width=3.5in]{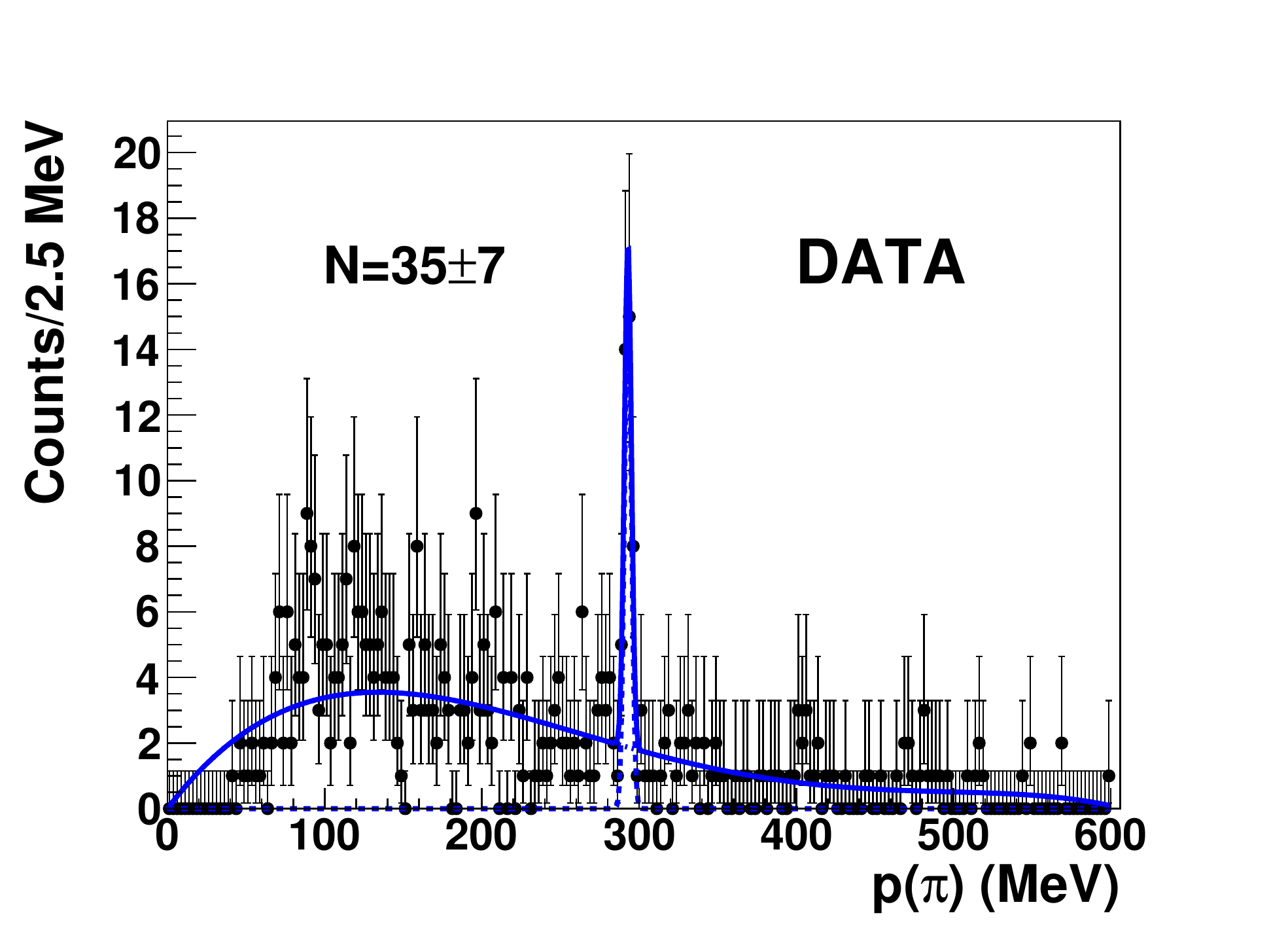}
\includegraphics[width=3.5in]{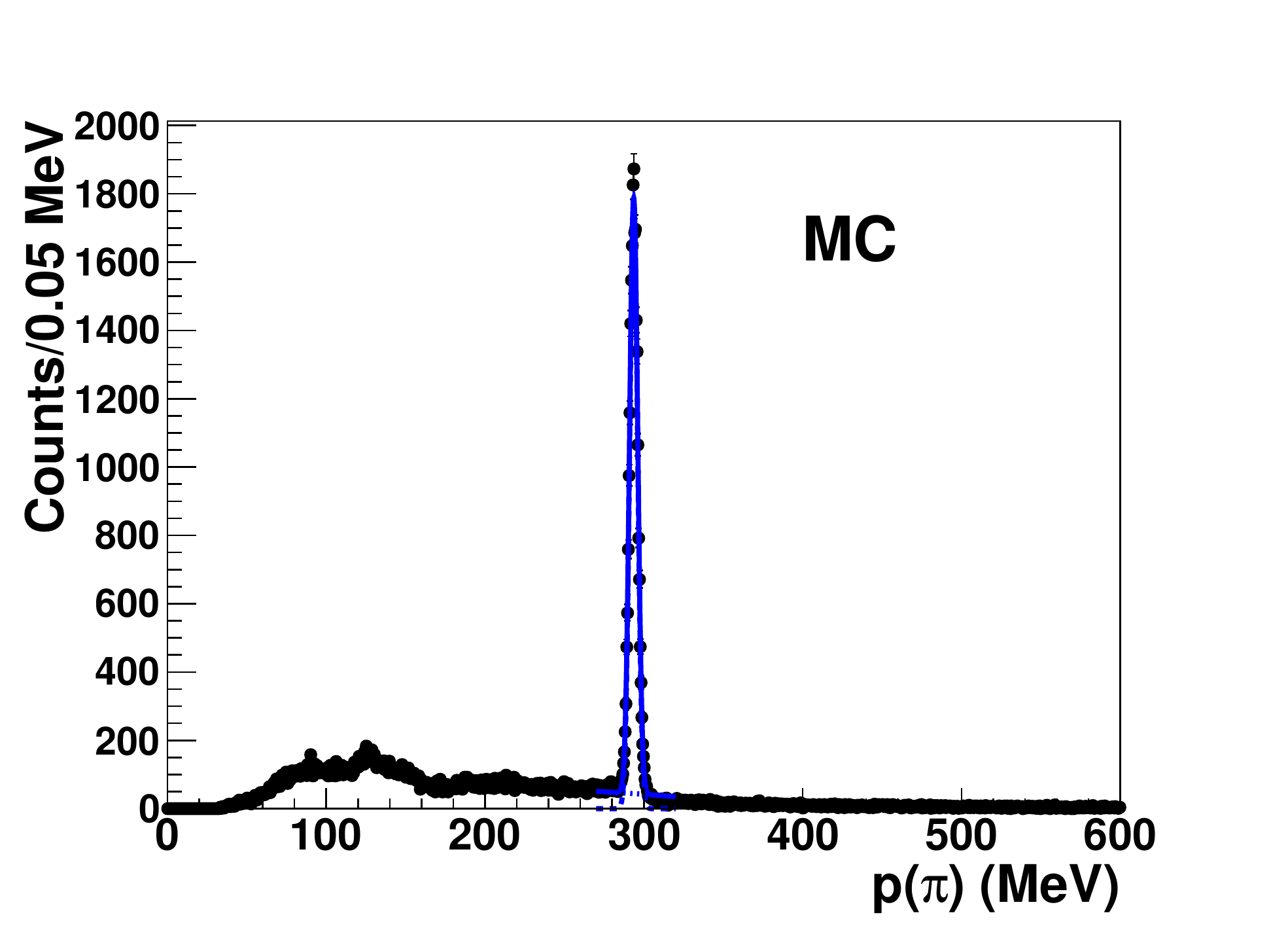}
\end{center}
\caption{Momentum distribution of tracks from $\Omega^-$ decays in the $\psi(2S)$ data (top) and MC (bottom), assuming they are pions. 
}
\label{fig:pi_mom}
\end{figure}

To measure the branching fraction for the channel $\Omega^- \to \Xi^0 \pi^-$, we reject kaon tracks that lie in the range of 195 to 230 MeV as shown in Fig.~\ref{fig:k_mom}. The momentum distribution of the remaining tracks in the rest frame of $\Omega^-$, assuming they are pions, is shown in Fig.~\ref{fig:pi_mom}. We fit the momentum spectrum with a Gaussian signal and a 3rd order polynomial background. Even though there is a big background in the low momentum region, the fit is able to identify the peak very well. The result from the fit gives $N_\pi = 35\pm7$.

The MC-determined efficiency for the detection of these transition pions, $\epsilon_\pi=53.3\%$. 
This leads to $\mathcal{B}(\Omega^- \to \Xi^0 \pi^-) = (35\pm7)/(0.511\times(360\pm23))=(19.0\pm4.0)\%$.

\begin{table}[!tb]
\caption{Summary of systematic uncertainties.  Sources $1-5$ are from~\cite{hyperon}.
The total systematic uncertainty listed in the sum in quadrature of the individual contributions.}
\begin{tabular}{l|c|c}
\hline
Sources              &  $\Omega^- \to \Lambda^0 K^- (\%)$  & $\Omega^- \to \Xi^0 \pi^- (\%)$\\
\hline 
$N(\Omega^-)$              &   6  & 6\\
$\Omega^-$  peak fitting        &   1 &       1 \\
Track reconstruction         & 1  & 1\\
$K^-$, $\pi^-$ peak fitting    &     1 &      2 \\
\hline
\textbf{~~~Total}                & \textbf{6.2} & \textbf{6.5}\\
\hline

\end{tabular}
\label{tbl:systematics}
\end{table}

\section{Systematic Uncertainties}

We evaluate systematic uncertainties due to various sources for each final state and add the contributions from the different sources together in quadrature. The uncertainty due to the number of $\Omega^-$ is 6\%. The uncertainty due to track reconstruction is 1\% per charged particle.  Uncertainties in $\Omega^-$ hyperon peak fitting and track momentum peak fitting are evaluated by varying the order of the polynomial background and the fit range.  The largest variations of these are taken as the estimates of systematic uncertainty in peak fittings.  The individual values and quadrature sums are given in Table~\ref{tbl:systematics}.

\section{Summary and Discussion of Results}
In summary, we have measured $\Omega^-$ decay branching fractions using CLEO $e^+e^-$ annihilation data. The results are summarized in Table~\ref{tbl:result}.

\begin{table}[h]
\caption{Summary of branching fraction results for the two body decays of $\Omega^-$ to $\Lambda^0 K^-$ and $\Xi^0 \pi^-$.}
\begin{tabular}{l|l|l}
\hline
               &  $\Omega^- \to \Lambda^0 K^- (\%)$  & $\Omega^- \to \Xi^0 \pi^- (\%)$\\
\hline 
CLEO DATA              &   $68.9\pm9.5\pm4.3$  & $19.0\pm4.0\pm1.2$\\
CERN SPS~\cite{omega}         & $67.8\pm0.7$ & $23.6\pm0.7$\\

\hline

\end{tabular}
\label{tbl:result}
\end{table}


\begin{acknowledgments}
We wish to thank Prof. Michael Schmitt for bringing the importance of the present independent measurement of $\Omega^-$ branching fractions to our attention.
This research was done using CLEO data, and as members of the former CLEO Collaboration we thank it for this privilege.  This research was supported by the U.S. Department of Energy, Office of Nuclear Physics, under Contract No. DE-FG02-87ER40344.
\end{acknowledgments}

\end{document}